\documentclass[preprint]{ptephy_v1}

\preprintnumber{XXXX-XXXX} 
\usepackage{hyperref}
\usepackage{dirtytalk}
\usepackage{wrapfig}
\usepackage{svg}
\usepackage{subcaption}
\usepackage{xspace}
\newcommand{\gammaray}{\ensuremath{\gamma}-ray\xspace}
\newcommand{\gammarays}{\ensuremath{\gamma}-rays\xspace}

\usepackage{lineno}

\begin{document}

\title{
Detectors for next-generation quasi-free scattering experiments at the RIBF
}


\author[1,2]{Junki Tanaka}
\affil[1]{RCNP, Osaka University, 10-1 Mihogaoka, Ibaraki, Osaka, Japan}
\affil[2]{RIKEN Nishina Center, 2-1 Hirosawa, Wako, Saitama 351-0198, Japan}

\author[2]{Martha Liliana Cort\'es} 

\author[3]{Hongna Liu} 
\affil[3]{Key Laboratory of Beam Technology of Ministry of Education, College of Nuclear Science and Technology,
Beijing Normal University, Beijing 100875, China}

\author[4]{Ryo Taniuchi} 
\affil[4]{School of Physics, Engineering and Technology, University of York, York, YO10 5DD, United Kingdom}


\begin{abstract}%
The advent of high-intensity radioactive ion beams has opened new avenues for nuclear structure research. 
By studying exotic ions, phenomena such as shell evolution, halos, and the limits of stability have been studied. 
In particular, Quasi-Free Scattering (QFS) experiments on hydrogen targets have proven to be a valuable tool to investigate the structure of exotic ions. 
Recently, a series of QFS experiments were performed at the Radioactive Isotope Beam Factory (RIBF) of the RIKEN Nishina Center employing the MINOS liquid hydrogen system. 
A fundamental part of the success of these experiments was the use of dedicated devices to measure the \gammarays, charged particles, and neutrons emitted in the reactions. 
The experience gained during the past campaigns, as well as the upcoming upgrade of the RIBF facility call for improvements on the existing devices, as well as for the development of new detection systems. 
Here we review the main detection devices used at the RIBF for QFS experiments, and give an overview of the ongoing and upcoming developments. 

\end{abstract}

\subjectindex{xxxx, xxx}

\maketitle

\section{Introduction}

Recent advances in the production of Radioactive Ion Beams (RIBs) have transformed modern nuclear physics by allowing us to study the structure of exotic nuclei. This has provided us essential tests for models of the nuclear force and of nuclear reactions under extreme conditions~\cite{Nowacki_PPNP_120_2021}.
Furthermore, precision measurements of nuclear masses, half-lives, and reaction cross sections have improved astrophysical models of supernovae and neutron star mergers~\cite{Aumann_PPNP_112_2020}.
At the Radioactive Isotope Beam Factory (RIBF) of the RIKEN Nishina Center~\cite{Yano_NIMB2007}, the unparalleled high primary beam intensities~\cite{Accelerator2021_RAPR,Accelerator2022_RAPR} have been exploited to study phenomena such as the limits of nuclear existence~\cite{Kondo_Nature_2023,Ahn_PhysRevLett.129.212502_2022}, shell evolution~\cite{Steppenbeck_Nature_2013,Taniuchi_Nature_2019},  and multi-nucleon resonances~\cite{Kisamori_PhysRevLett_116_2016,Duer_Nature_2022}. 

\begin{figure}[bth]
\centering
\includegraphics[width=0.9\textwidth]{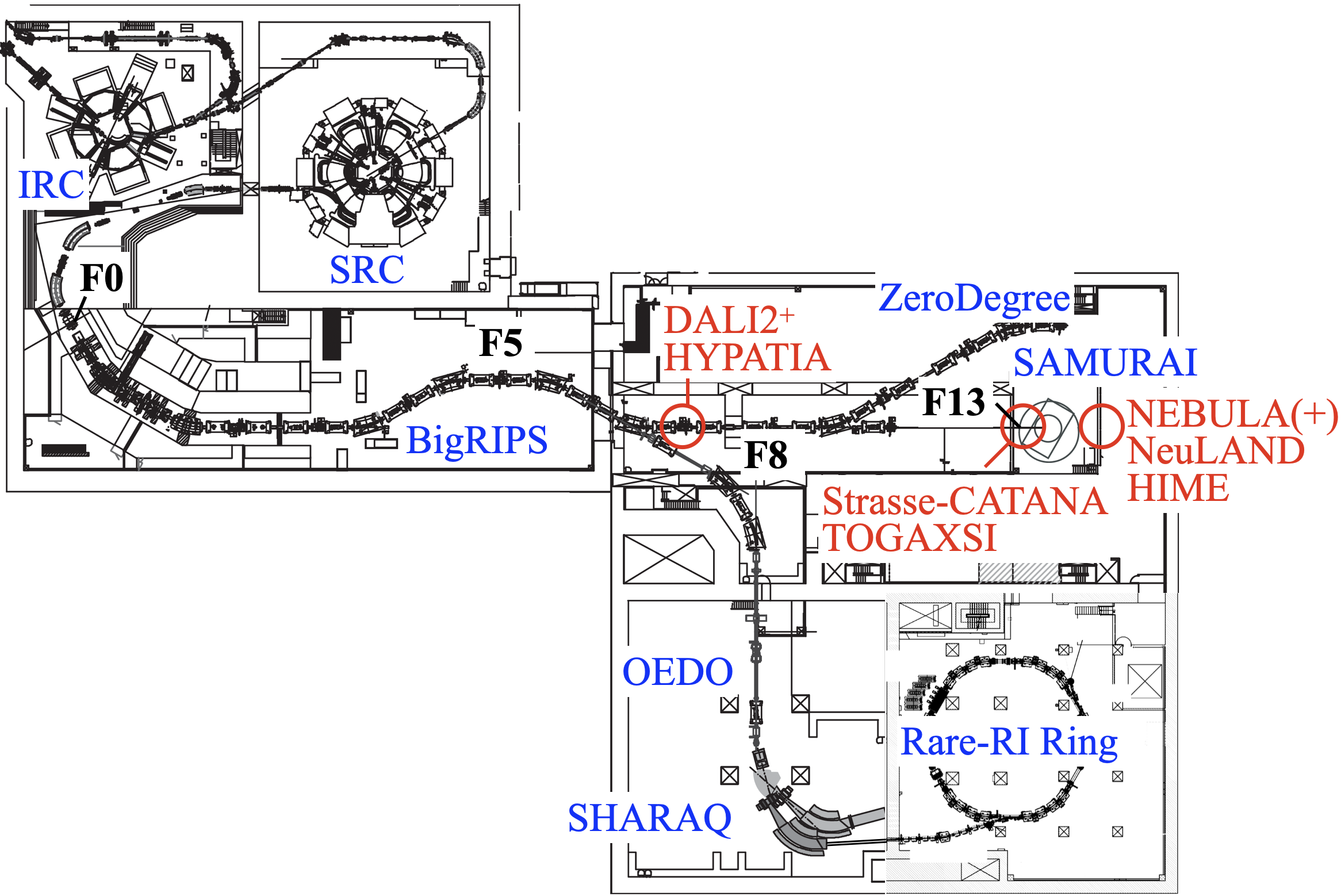}
\caption{Schematic of the RIBF facility. After the SRC, stable beam are delivered to the BigRIPS fragment separator where RIBs are produced, separated and identified. The different spectrometers available at the RIBF are shown, and different devices used for quasi-free scattering experiments are listed in red, at the focal plane where they are placed.}
\label{fig:ribf}
\end{figure}

Figure \ref{fig:ribf} shows the layout of the RIBF.
Stable beams are accelerated up to 345~MeV/u using four cyclotrons. The last two, the Intermediate-stage Ring Cyclotron (IRC), and the Superconducting Ring Cyclotron (SRC), can be seen in Fig.~\ref{fig:ribf}.
After exiting the SRC, beams are delivered to the F0 focal plane of the BigRIPS fragment separator~\cite{Kubo_PTEP_2012} where RIBs are produced via in-flight fragmentation or fission on a thick target. 
Reaction products are then separated and identified event-by-event based on their mass-to-charge ratio, $A/Q$, and atomic number, $Z$~\cite{Fukuda_NIMA2013}. 

The resulting secondary beams, with energies from a few tens to several hundreds MeV/u, are delivered to one of the spectrometers of the RIBF: ZeroDegree~\cite{Kubo_PTEP_2012}, SAMURAI~\cite{Kobayashi_NIM_317_2013}, SHARAQ~\cite{Michimasa_NIMA_317_2013}, OEDO~\cite{Michimasa_PTEP_2019}, and the Rare-RI Ring~\cite{YAMAGUCHI2013}, where different types of measurement can be performed. 
In particular, Quasi-Free Scattering (QFS) experiments have been mainly performed at the ZeroDegree and SAMURAI spectrometers.
In these measurements RIBs with energies around 200~MeV/u impinge on a secondary target, leading to the knockout of one or few  nucleons  and approximating a free nucleon-nucleon collision. 
With this  technique bound and unbound excited states, single-particle properties, and correlations within nuclei can be probed. 

In recent years, a series of experiments involving QFS were performed at the RIBF employing the MINOS liquid hydrogen target~\cite{Obertelli_EPJA_2014,santamaria2018tracking}. Thanks to the time projection chamber (TPC) surrounding the target, it was possible to deduce the interaction vertex with a resolution of about 5~mm (FWHM) which, in turn, allowed for the use of a thicker target with higher luminosity. Different experiments with target lengths varying between 10 and 15~cm were performed at F8 and F13, and different detection systems were employed to measure the reaction residues. 

At F8, the main focus was in-beam \gammaray spectroscopy following knockout reactions. In this case the DALI2$^+$ array~\cite{takeuchi2014dali2,Murray_RAPR_2018} was employed. This array, consisting of 226 NaI(Tl) detectors, provides high \gammaray detection efficiency which enables the study of bound excited states, as well as exclusive cross sections. 
At F13, MINOS was followed by the SAMURAI spectrometer to perform invariant mass measurements. In this case, in addition to the standard detectors of SAMURAI, which measure the $A/Q$ and $Z$ of the outgoing fragments, the neutron detectors NeuLAND~\cite{Boretzky_NIM_2021} and NEBULA~\cite{Kondo_NIM_2020} were employed. Furthermore, the DALI2$^+$ array was also used in some experiments at F13 to measure excited states. 

The combination of MINOS with these different detection systems was successfully used to measure phenomena such as magicity around $^{78}$Ni~\cite{Taniuchi_PTEP_2025}, shell evolution around the Ca isotopes~\cite{Lui_PTEP_2025} and the $N=40$ Island of Inversion (IoI)~\cite{Cortes_PTEP_2025},  the limits of the $N=20$ IoI~\cite{Kahlbow_PTEP_2025}, and the deformation of isotopes between Zn and Zr~\cite{Chen_PTEP_2025}.  
These measurements also highlighted possible improvements to the detections systems and provided new ideas to further optimize the measurements. The continuous developments on the instrumentation is, therefore, of fundamental importance for studies involving QFS. This is particular the case at the RIBF, where an upgrade plan of the facility has been outlined for the coming decade~\cite{ribf_upgrade}.

In this article we review the main detection systems used for QFS experiments at the RIBF and give an overview of the on-going and future developments. Section~\ref{sec:charged} focuses on charged particle detectors for missing mass measurements, in particular STRASSE and TOGAXI.  Section~\ref{sec:gamma} gives an overview on the developments  for  in-beam \gammaray spectroscopy such as the HYPATIA array. Section~\ref{sec:neutrons} focuses on neutron detection arrays, in particular HIME and NEBULA+.  In Sec.~\ref{sec:conclusion} a summary and outlook are presented.

\section{Missing-mass spectroscopy in inverse kinematics }\label{sec:charged}

The nucleons in the atomic nucleus can be considered to be a system of independent particles within the Fermi gas model.
The Fermi wave number, which defines the momentum of the highest occupied state in nuclear matter, is related to the typical nuclear saturation density of $\rho_0 \approx 0.17~\mathrm{fm}^{-3}$. 
This value translates to $k_F = [3\pi^2/(2\rho_0)]^{1/3}= 1.36~\textrm{fm}^{-1}$, or approximately 270~MeV/$c$ in the momentum space. 
When the momenta of the particles significantly exceed Fermi momentum, the particles enter a quasi-free condition, no longer bound by the nuclear force. 
Therefore, proton-induced knockout reactions in inverse kinematics, expressed as $(p,pX)$, at beam energies above 200~MeV/u can be considered QFS reactions, where the proton scatters off a quasi-free cluster or nucleon inside the target nucleus.

\begin{figure}[hbtp]
\centering
\begin{minipage}[c]{0.50\linewidth}
    \subcaptionbox{}[1.0\linewidth]{%
        \includegraphics[width=1.0\linewidth]{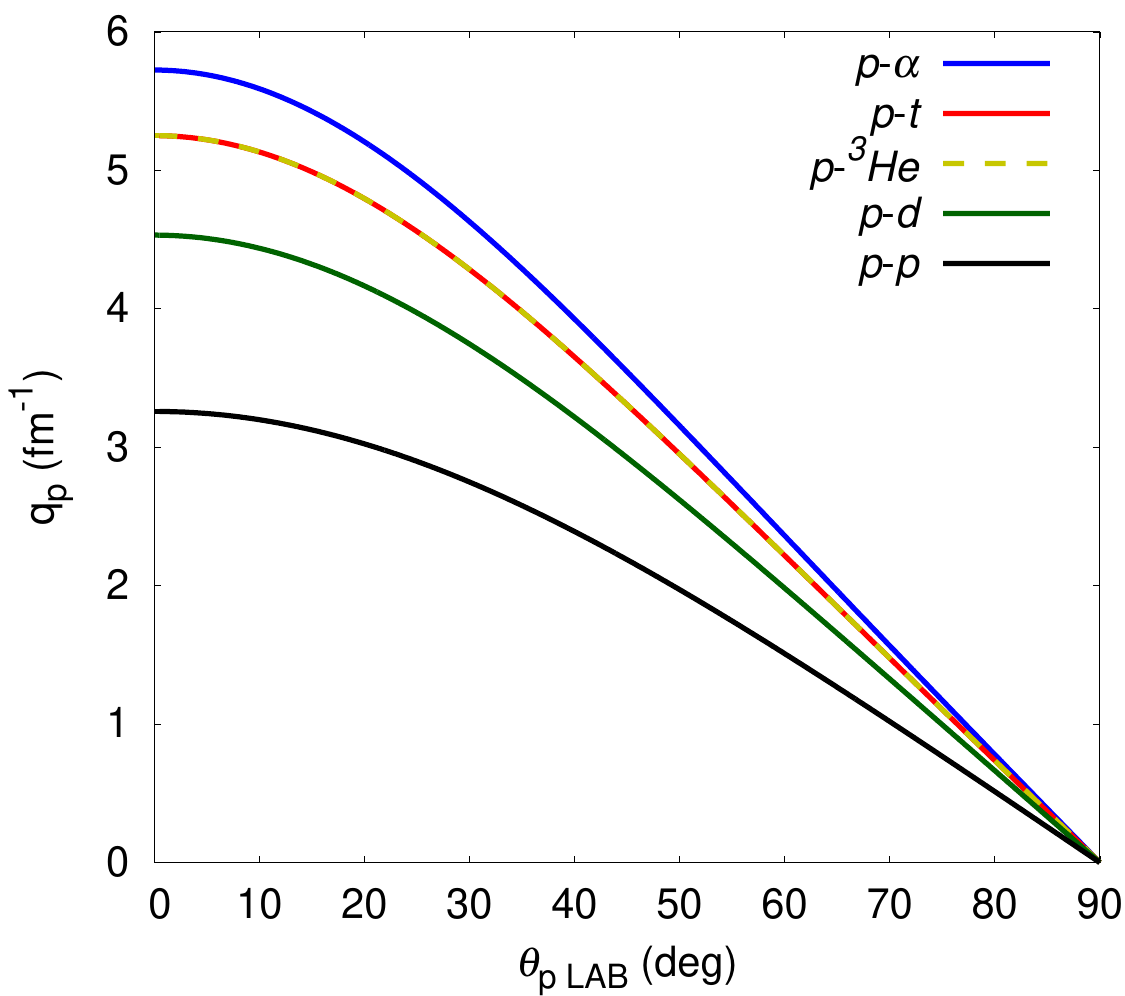}
    }
\end{minipage}%
\begin{minipage}[c]{0.50\linewidth}
    \subcaptionbox{}[1.0\linewidth]{%
        \includegraphics[width=1.0\linewidth]{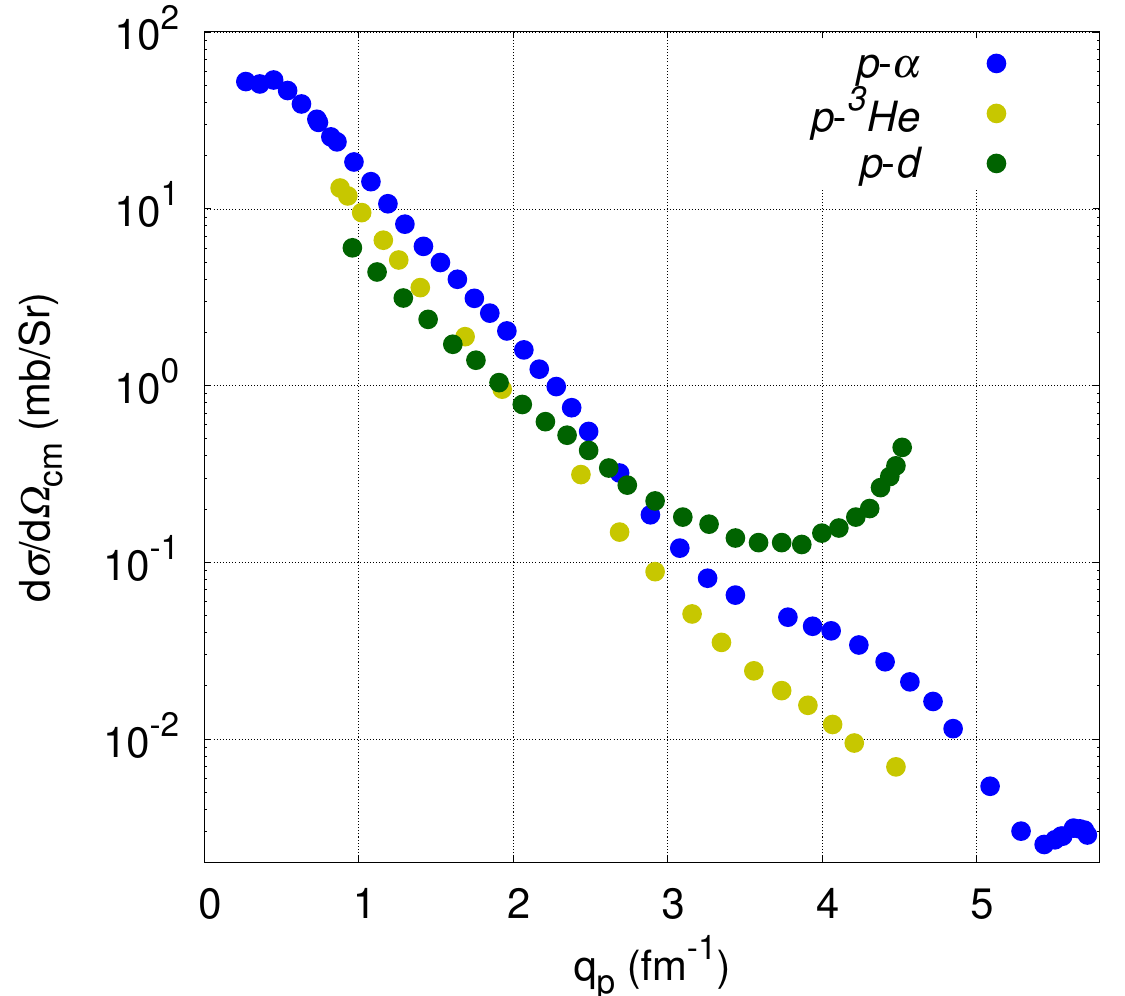}
    }
\end{minipage}
\vspace{-0.5cm}
\caption{(a) Dependence of recoil proton momentum on the scattering angle for different knockout reactions at 200 MeV/u.
(b) Dependence of the QFS differential cross section on the momentum transfer~\cite{ermisch2005systematic,hasell1986elastic,moss1980proton}.}
\label{fig:theta_momtrans}
\end{figure}

In $(p,pX)$ reactions the proton, initially at rest in the laboratory frame, gains momentum due to the reaction with the incoming heavy ion. Under the QFS approximation, it is assumed that the knocked-out particle acquires the same momentum transfer relative to the residual nucleus, as if the reaction was a direct, two-body collision, largely unaffected by the rest of the nucleus. 
Figure ~\ref{fig:theta_momtrans}a) illustrates the dependence of the momentum transferred to the recoil proton as a function of the scattering angle. 
As the transferred momentum increases, the impact parameter between protons and intra-nuclear particles before the knockout becomes smaller. Consequently, the quasi-free scattering cross-sections tend to decrease, as shown in Fig.~\ref{fig:theta_momtrans}b).

Missing-mass spectroscopy following QFS reactions in inverse kinematics has become a key technique for investigating exotic nuclear structures, particularly in weakly bound and unbound nuclei, clustered systems, or exotic decays where direct detection is impractical.
In this technique, undetected particles are identified by applying conservation of energy and momentum. 
Instead of detecting all final-state particles directly, the momenta and energies of the incoming beam and some of the outgoing particles are measured. The mass of the unobserved particle(s) is then deduced as the “missing” component needed to balance the reaction kinematics.

In particular for $(p,pX)$ reactions, the missing mass, $M_{\text{miss}}$, is deduced as

\begin{align}
M_{\text{miss}}=\sqrt{\left\{E_b+m_p-(E_p+E_X)\right\}^2-\left\{\vec{P_b}-(\vec{P_p}+\vec{P_X})\right\}^2},
\end{align}
where $E_i$, and $\vec{P_i}$ represent the energy and the three-momentum vector of the beam ($i=b$), the recoil proton ($i=p$) and the knocked-out particle ($i=X$), respectively, and $m_p$ is the proton mass. 

\begin{figure}[hbtp]
\centering
\begin{minipage}[c]{0.49\linewidth}
    \subcaptionbox{}[1.0\linewidth]{%
\includegraphics[width=1.0\hsize]{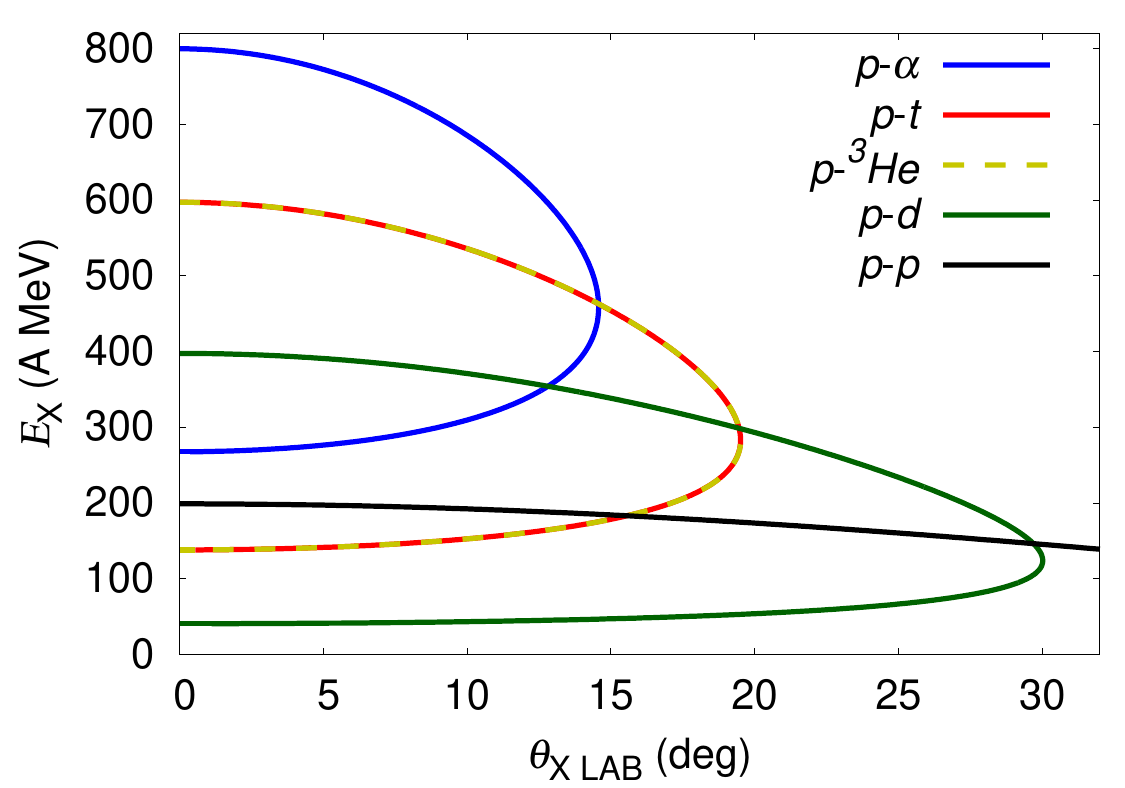}
}
\end{minipage}\ 
\begin{minipage}[c]{0.49\linewidth}
    \subcaptionbox{}[1.0\linewidth]{%
\includegraphics[width=1.0\hsize]{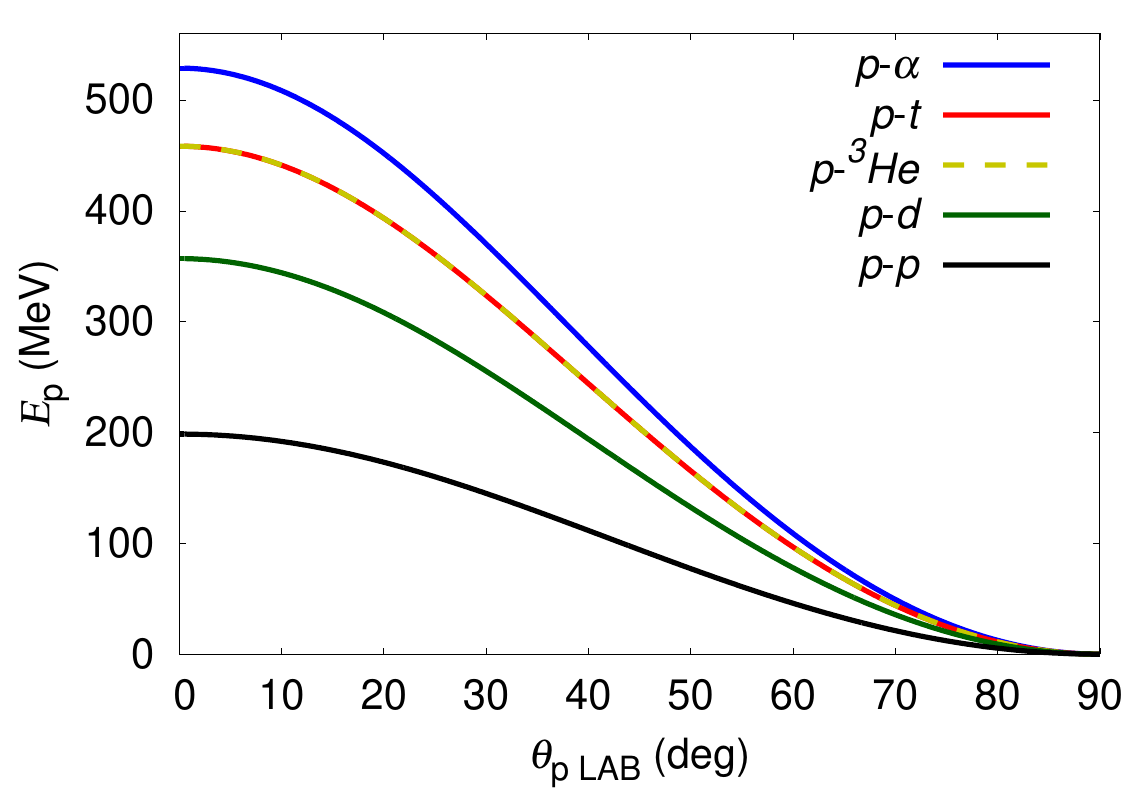}
}
\end{minipage} 
\vspace{-0.5cm}
\caption{Relation between emission angle and energy of (a) the scattered particles X, and (b) the recoil protons in  $(p,pX)$ reactions at 200 MeV/u. The reaction Q-values are assumed to be zero for these calculations.}
\label{kinematics3}
\end{figure}

Figure~\ref{kinematics3} shows the relation between the energy, and emission angle, $\theta$, of a) the knocked-out particles (such as protons, deuterons, $^3$He, tritons, and $\alpha$ particles), and b) the recoil protons, for $(p,pX)$ reactions at incident energy of 200~MeV/u. For simplicity, the internal momentum of the knocked-out particle in the initial state is not taken into account.
It can be seen that the mass of the knocked-out particles significantly influences $\theta_X$, particularly the maximum angles in which detection is possible: While knocked out protons can be detected at a larger angular range, heavier ejectiles are only emitted at forward angles, smaller than 30$^{\circ}$. In contrast, as can be seen on Fig.~\ref{kinematics3}b), the recoil protons are scattered up to 90$^{\circ}$, regardless of the knocked-out particle. 

When designing instrumentation for QFS reactions, it is essential to consider the angular distributions and energies of the scattered particles, as well as the cross sections required to achieve realistic yields. Detectors with different characteristics are therefore necessary to address different physics questions that can be studied with QFS reactions. 
The next subsections describe two major ongoing developments: STRASSE, optimized for nucleon knockout, and TOGAXSI, designed to study cluster knockout reactions. 
Each system is introduced with its specific motivation and improvement over existing approaches.




\subsection{STRASSE: A telescope array for $(p,2p)$ and $(p,3p)$ reactions}

STRASSE (Silicon Tracker for RAdioactive nuclei Studies at SAMURAI Experiments)~\cite{strasse_epja} is a detection system optimized for $(p,2p)$ and $(p,3p)$ reactions at 200--250~MeV/nucleon. 
Similarly to MINOS, a thick LH$_2$ target with a length up to 150 mm is used to induce the QFS reactions. The recoil protons are detected, not by a TPC, but by a highly segmented Si tracker array placed in vacuum to minimize the angular straggling of protons. With this setup, proton detection efficiencies of 86\% and 49\% for one and two protons, respectively, can be achieved, comparable to those obtained with MINOS. 
Thanks to the high segmentation of the Si tracker, a vertex resolution of 0.5 mm (FWHM) and an angular resolution of $\sim12$ mrad (FWHM), can be obtained. 

STRASSE will be coupled with the CATANA (Cesium iodide Array for $\gamma$-ray Transitions in Atomic Nuclei at high isospin Asymmetry) array~\cite{catana}, composed of 140 CsI(Na) scintillators arranged into 7 rings (L2-L8), and covering  polar angles from 17 to 77$^{\circ}$. CATANA crystals can stop protons up to 250 MeV. In particular, a dual-gain readout has been developed for CATANA to enable simultaneous \gammaray and proton measurement. Note that the 20 crystals in L2 will be used to detect only protons. With the STRASSE + CATANA setup,  missing-mass spectra, which could not be achieved with the MINOS system, are expected to be available with a resolution better than 4~MeV (FWHM). This offers an excellent opportunity to study highly unbound states (such as 4n or 6n states) or isomeric bound states (such as the $0_2^+$ states of doubly magic nuclei) at the RIBF.

\begin{figure}[bth]
\begin{center}
    \includegraphics[width=0.65\textwidth]{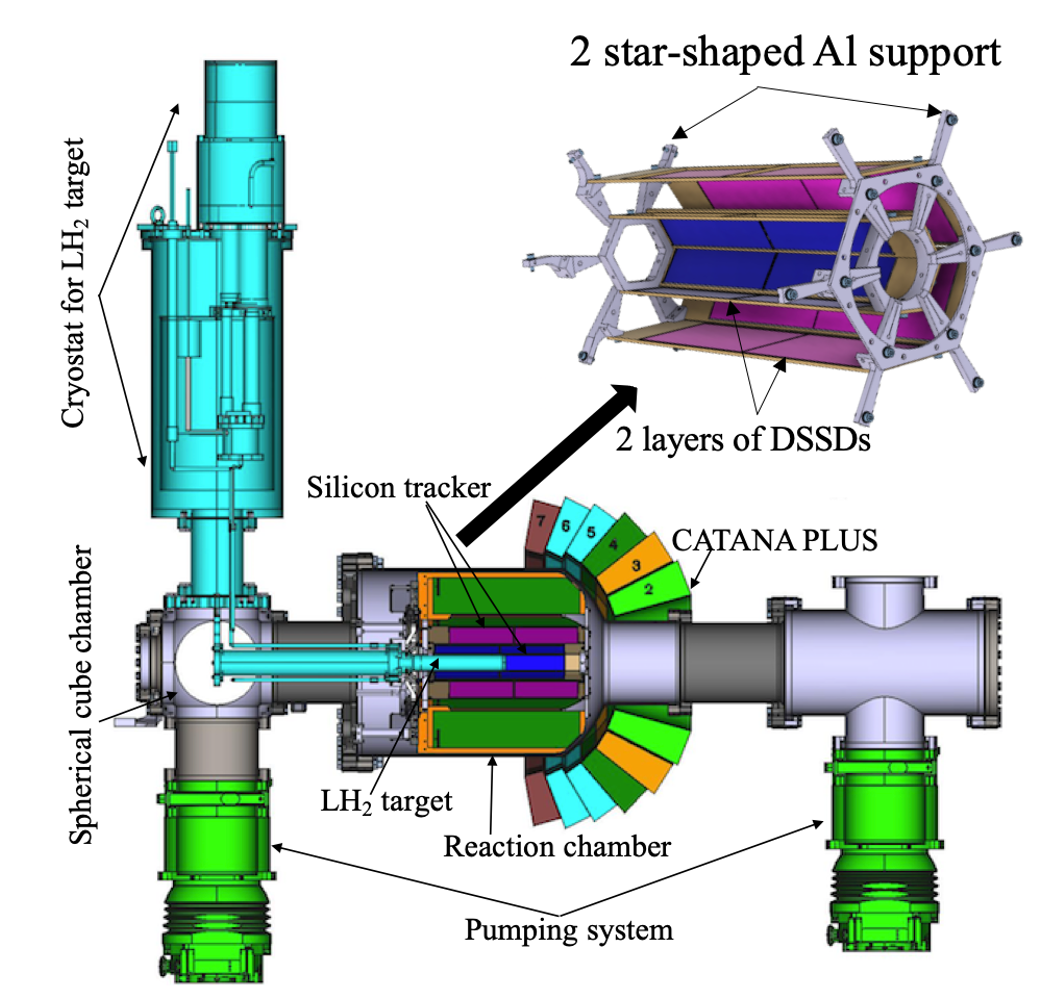}
    \end{center}
    \caption{Schematic view of the overall conceptual design of the STRASSE detection system.}
    \label{fig:strasse}
\end{figure}

The schematic view of the setup, currently under construction, can be seen in Fig.~\ref{fig:strasse}.
The geometry of the LH$_2$ target in STRASSE is designed to minimize the energy loss and angular straggling of the protons produced from QFS. The target system consists of three parts: 1)~the target cell made of 175 $\mu$m Mylar foil with a 31-mm diameter cylinder and a 20-mm diameter entrance window (the difference is due to the hydrogen inlet and outlet pipes), 2)~the cryostat to liquefy the hydrogen at $\sim$20~K, and 3)~the control command system and the gas circuit to ensure the safe operation of the target.
The hydrogen circulation in the STRASSE target is driven by the thermal siphon as the former PreSPEC~\cite{prespec} and MINOS~\cite{Obertelli_EPJA_2014} targets. As shown in Fig.~\ref{fig:strasse}, two pumping stations located upstream and downstream of the STRASSE reaction chamber will be used to achieve the required vacuum. STRASSE will be a standalone system with windows to isolate from the beamline pipes. 

The STRASSE silicon tracker is composed of two layers of double-sided silicon strip detectors (DSSDs) configured in a hexagonal shape, as shown in Fig.~\ref{fig:strasse}. 
Each segment includes two DSSDs with strips parallel to the beam axis daisy chained. The inner segment has an active area of 30~mm $\times$ 244~mm with a thickness of 200~$\mu$m and a pitch size of 200~$\mu$m, while the outer segment has an active area of 62.6~mm $\times$ 242~mm with a thickness of 300~$\mu$m and a pitch size of 200~$\mu$m. 
The whole silicon tracker has a total of 17,358 electronic channels. 128-channel STS-XYTER~\cite{sts-xyter} ASIC chips with the ability to detect minimum ionizing particles are used to read out the signals from DSSDs via custom-made low-mass and low-capacitance microcables. It features a 14-bit time stamp and a 5-bit flash ADC. The STS-XYTER chips are wire-bonded to the front-end PCBs (FEBs) placed in the radial direction inside the vacuum chamber to achieve the required equivalent noise charge of $\sim$10~keV. A cooling system is designed to dissipate the heat load of the electronics. Two star-shaped stainless steel frames support the detectors. The data from different FEBs are then sent to a concentration board called GBTxEMU via a mezzanine connection board. A PCI-express GERI board receives the aggregated data via optical fibers and synchronizes different GBTxEMUs.


\subsection{TOGAXI: Elucidating cluster formation with $(p,pX)$ telescope}

Clustering is a fundamental feature of nuclear structure, where nucleons tend to group into substructures such as $\alpha$ particles within a nucleus. By measuring cluster knockout reactions, it is possible to investigate the existence of clusters on the surface and inside of atomic nuclei.
A notable example of $\alpha$-cluster formation in heavy ions was recently reported for the stable Sn isotopes~\cite{Tanaka_Science_2021}. 
Analyzing the emitted particles after the $(p,p\alpha)$ reaction using a magnetic spectrometer, the cluster separation energy, a characteristic physical quantity of cluster knockout reaction, was deduced.
Based on its agreement with theoretical models, a close connection between the $\alpha$-cluster formation and the neutron skin was suggested.

\begin{figure}[tb]
\begin{center}
    \includegraphics[width=0.65\textwidth]{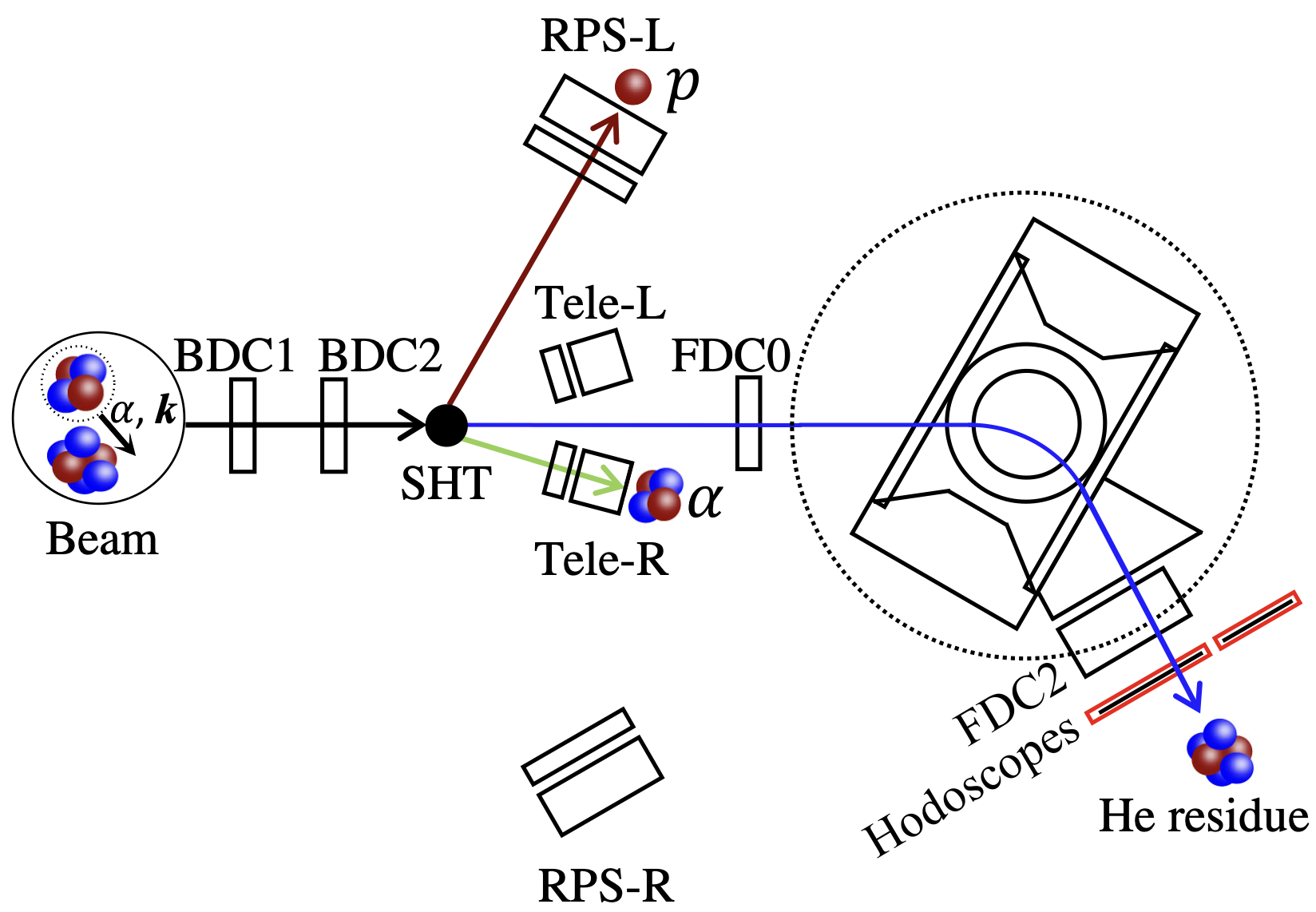}
    \end{center}
    \caption{Experimental setup at SAMURAI experimental area for $^{10}$Be$(p,p\alpha)$ reaction. The figure is taken from~\cite{li2023validation}.}
   \label{fig:10Be}
\end{figure}

Similar types of studies have also been performed  for exotic isotopes by using liquid or solid hydrogen targets and inverse kinematics.
For example, in a recent measurement at the SAMURAI spectrometer, the measured triple differential cross section as a function of the proton energy revealed the existence of $\alpha$-clusters in $^{10}$Be, in agreement with a 2$\alpha$-2$n$ configuration~\cite{matsuda2011large}. 
Figure~\ref{fig:10Be} shows the experimental setup used in that study.
A 2-mm-thick solid hydrogen target (SHT)~\cite{matsuda2011large} was employed to induce $(p,p\alpha)$ reactions. The $^6$He residual nuclei were analyzed using the forward-located SAMURAI spectrometer.
Because the emitted protons and alpha particles differ significantly in both emission angle and energy, separate detectors were placed to match each particle's emission characteristics.
Protons emitted at 60 degrees relative to the beam axis were measured with a recoil proton spectrometer (RPS-L(R)), composed of a drift chamber and a NaI (Tl) scintillator~\cite{chebotaryov2018proton,matsuda2013elastic}, and $\alpha$ particles emitted between 4 and 12 degrees, were detected using a telescope consisting of a silicon detector and a CsI (Tl) scintillator (Tele-L(R))~\cite{verde2013farcos}.
By placing the detectors at a relatively large distance from the target, which minimizes the relative opening angle, an $\alpha$ separation energy resolution of 2.5~MeV (FHWM) was achieved.
 In addition, the time of flight of the emitted protons was measured. 

\begin{figure}[tb]
\begin{center}
    \includegraphics[width=0.4\textwidth]{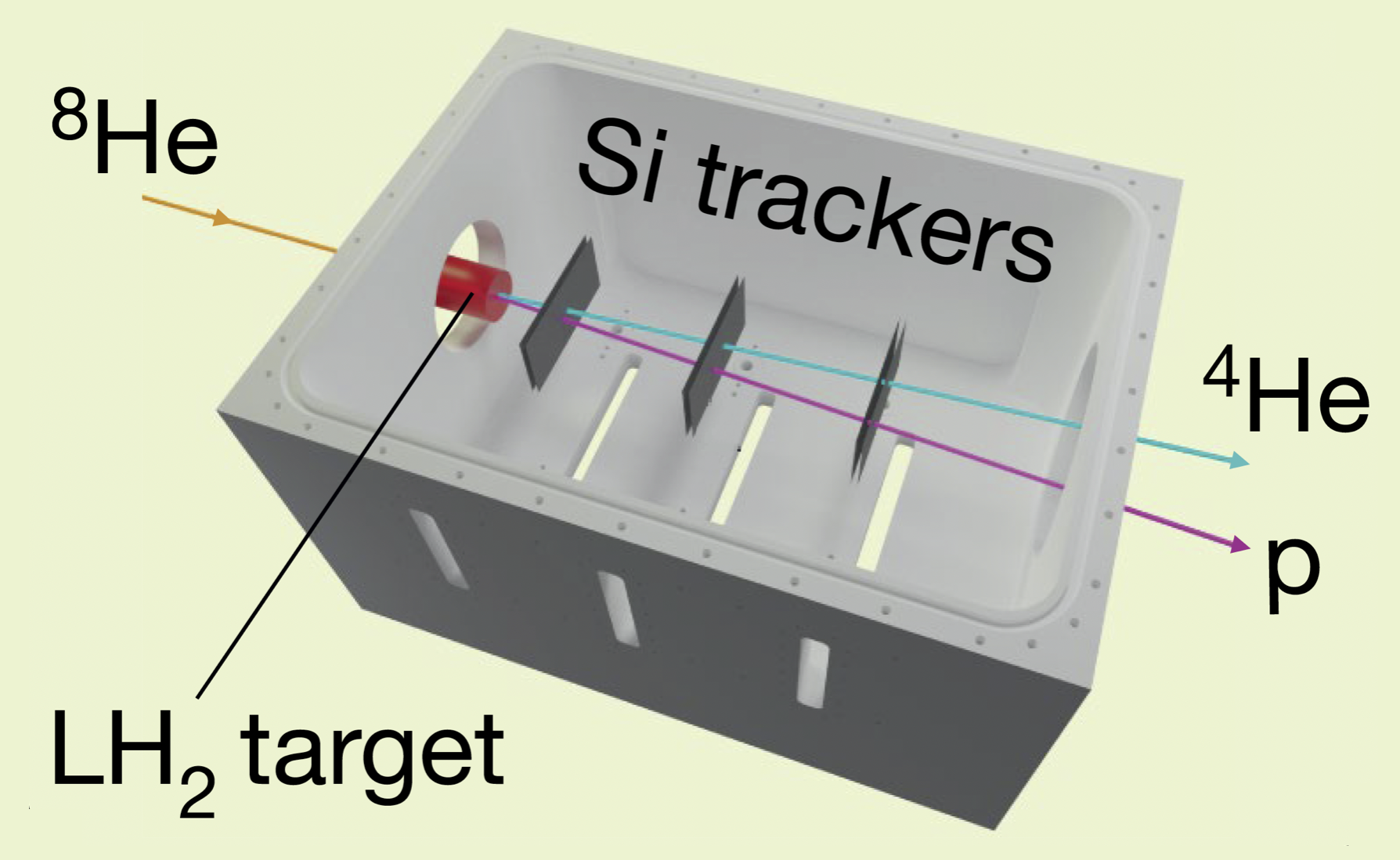}
    \end{center}
    \caption{Silicon trackers for $^8$He$(p,p\alpha)$ reaction experiment. The figure is taken from~\cite{Duer_Nature_2022}.}
    \label{fig:tetra}
\end{figure}

Alpha knockout reactions can also be used as a tool to explore loosely bound systems. 
A experiment searching for the tetra-neutron state was performed at the SAMURAI spectrometer using the $^8$He$(p, p\alpha)$ reaction.
In this case, silicon-strip detectors were placed downstream of a 50-mm-thick LH$_2$ target, as shown in Fig.~\ref{fig:tetra}. The detector configuration was optimized for events in which $\alpha$ clusters in $^8$He nuclei interact with protons at small impact parameters. Forward-emitted protons and recoil $\alpha$ particles were tracked using three layers of silicon-strip detectors, and their momenta were analyzed with the SAMURAI spectrometer.

In the cases described above, missing mass energy resolutions as low as 2.4~MeV (FWHM) has been achieved; however, the limited solid angle coverage remains a point for improvement. For next-generation detectors, it will be essential to inherit the successful concepts demonstrated by MINOS -- namely, the use of a thick target and a detection system with large solid-angle coverage -- to compensate for the low intensity of RI beams and to achieve high detection efficiency. Ideally, the system should be capable of measuring multiple types of cluster knockout reactions over a wide solid angle within a single configuration.
Incorporating a total energy calorimeter for the emitted particles can further ensure accurate missing mass spectroscopy. 
The detector array introduced in the following paragraph is designed to simultaneously detect these charged particles, enabling a missing mass energy resolution of approximately 3 MeV (FWHM).  

\begin{figure}[hbtp]
\begin{center}
    \includegraphics[width=0.5\textwidth]{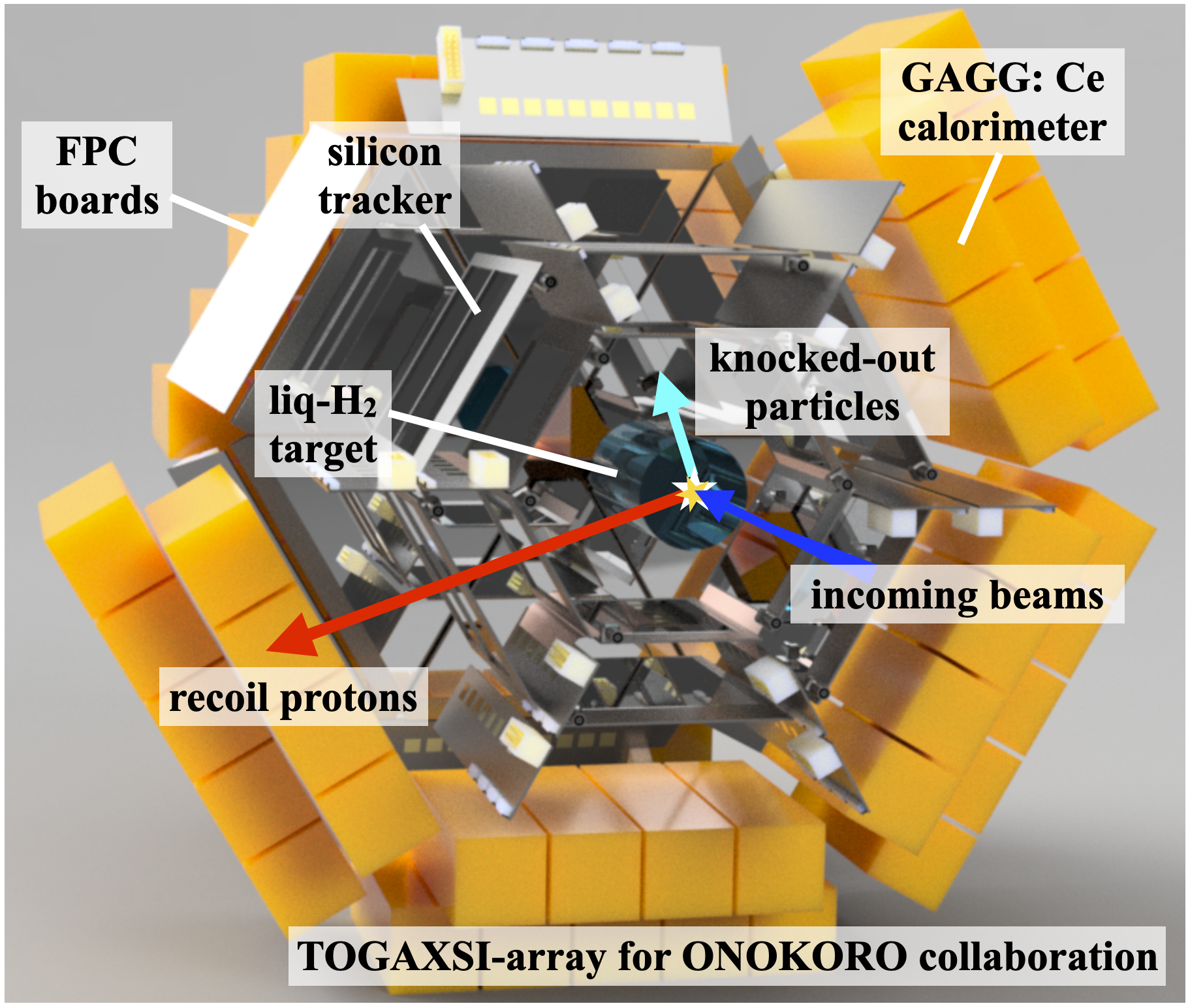}
    \end{center}
    \caption{Graphical design of TOGAXSI array. Incoming beams interact in the LH$_2$ target to produce cluster-knockout reactions. A Si tracker and a GAGG:Ce calorimeter are employed to measure the scattered clusters and the recoil protons and performe missing mass measurements. }
    \label{fig:onokoro}
\end{figure}

Figure~\ref{fig:onokoro} shows the TOGAXSI (TOtal energy measurement by GAgg and verteX measurement by SIlicon strips) telescope~\cite{TANAKA20234}, being developed  within the ``ONOKORO'' project~\cite{uesaka2021comprehensive} for dedicated measurements of cluster knockout reaction at the RIBF. 
Similarly to MINOS, it uses a thick LH$_2$ target which provides high luminosity. For ONOKORO, the same type of liquid hydrogen target used for the STRASSE has been adopted.
The target is surrounded by a Si tracker for vertex reconstruction and a GAGG:Ce calorimeter for energy measurement, which ensures accurate missing mass spectroscopy. 

The silicon tracker consists of three layers of silicon-strip detectors with a thickness of 100~$\mu$m and a pitch of 100~$\mu$m~\cite{HIGUCHI202384} which provide a  3~mrad angular tracking resolution (FWHM) and 0.5~mm (FHWM) vertex reconstruction resolution. 
GAGG(Ce) has been adopted as the material of the calorimeter due to its high density of 6.63~g/cm$^3$, which results in a high particle stopping power. Furthermore, it provides a high energy resolution of 2~MeV(FWHM) and is a non-deliquescent scintillation crystal. Relatively large crystals of $35~\textrm{mm} \times 35~\textrm{mm} \times 120~\textrm{mm}$ were chosen to stop high-energy charged particles. 

As can be seen in Fig.~\ref{kinematics3}a), under QFS conditions in inverse kinematics, the emitted clusters have high energies and are directed towards relatively forward angles, below 30$^{\circ}$, whereas recoil protons are emitted at larger angles. 
Consequently, the detector is designed to detect clusters at polar angles of $8^\circ$--$30^\circ$.
For the detection of recoil protons, detection angles between $35^\circ$--$70^\circ$ are selected. As can be seen in Fig.~\ref{fig:theta_momtrans}, for these angles a momentum transfer between $2$--$3$~fm$^{-1}$ is expected. This in turn leads to cross sections from $10^{-1}$ to $10^{-2}$~mb/Sr, which allows to keep a realistic yield for the experiments. 
In these angles the expected energy of the protons varies from $80$--$250$~MeV, and the cluster energy is between $75$--$180$~MeV/u.

The construction of the full array is underway. TOGAXSI is expected to accelerate comprehensive cluster research with missing mass spectroscopy based on the detection of charged particles. Table~\ref{tab:detector_comparison} summarizes the properties of MINOS, STRASSE and TOGAXI.

\begin{table}[btp]
\centering
\caption{Comparison of inverse-kinematics proton-induced reaction detector systems at RIBF. }
\label{tab:detector_comparison}
\begin{tabular}{lccc}
\hline
\textbf{Parameter} & \textbf{MINOS}~\cite{Obertelli_EPJA_2014} & \textbf{STRASSE}~\cite{strasse_epja} & \textbf{TOGAXSI}~\cite{TANAKA20234} \\
\hline
Vertex resolution (FWHM) & $\sim5$ mm  & $\sim0.5$ mm  & $\sim0.5$ mm  \\
Angular resolution (FWHM) & $\sim38$ mrad & $\sim12$ mrad & $\sim10$ mrad  \\
Missing-mass resolution (FWHM)&  ---  & $\leq 4$ MeV  & $\sim3$ MeV  \\
Angular coverage (cluster) & ---  & --- & $8^\circ$--$ 30^\circ$ \\
Angular coverage (recoil $p$) &  $14^\circ$--$ 70^\circ$ & $14^\circ$--$ 70^\circ$ & $35^\circ$--$ 70^\circ$ \\
\hline
\end{tabular}
\end{table}

\section{Gamma-ray detection for gamma-ray spectroscopy}\label{sec:gamma}

In-beam $\gamma$-ray spectroscopy is a powerful tool that allows the study of nuclear structure and excited states of atomic nuclei.  The advent of RIBs has increased the potential of this technique as new mass regions can be explored and new phenomena can be observed.  Indeed, high primary beam intensities have enabled the production of short-lived isotopes which in turn can impinge on solid or liquid targets to induce Coulomb excitation, inelastic scattering, knockout, and two-step fragmentation experiments in inverse kinematics~\cite{Doornenbal_PTEP_2012}. 

At the RIBF, secondary beams with  incident energies of 100 to 250~MeV$/u$ are used to populate excited states of nuclei far from stability.
While such energies allow the use of thicker targets, and therefore increase the luminosity, they also present different experimental challenges. Of particular importance are the large Doppler shift of the detected \gammarays, and the background generated by either atomic or nuclear processes. 

\begin{figure}[bt]
    \centering
    \includegraphics[angle=-90,width=0.6\linewidth]{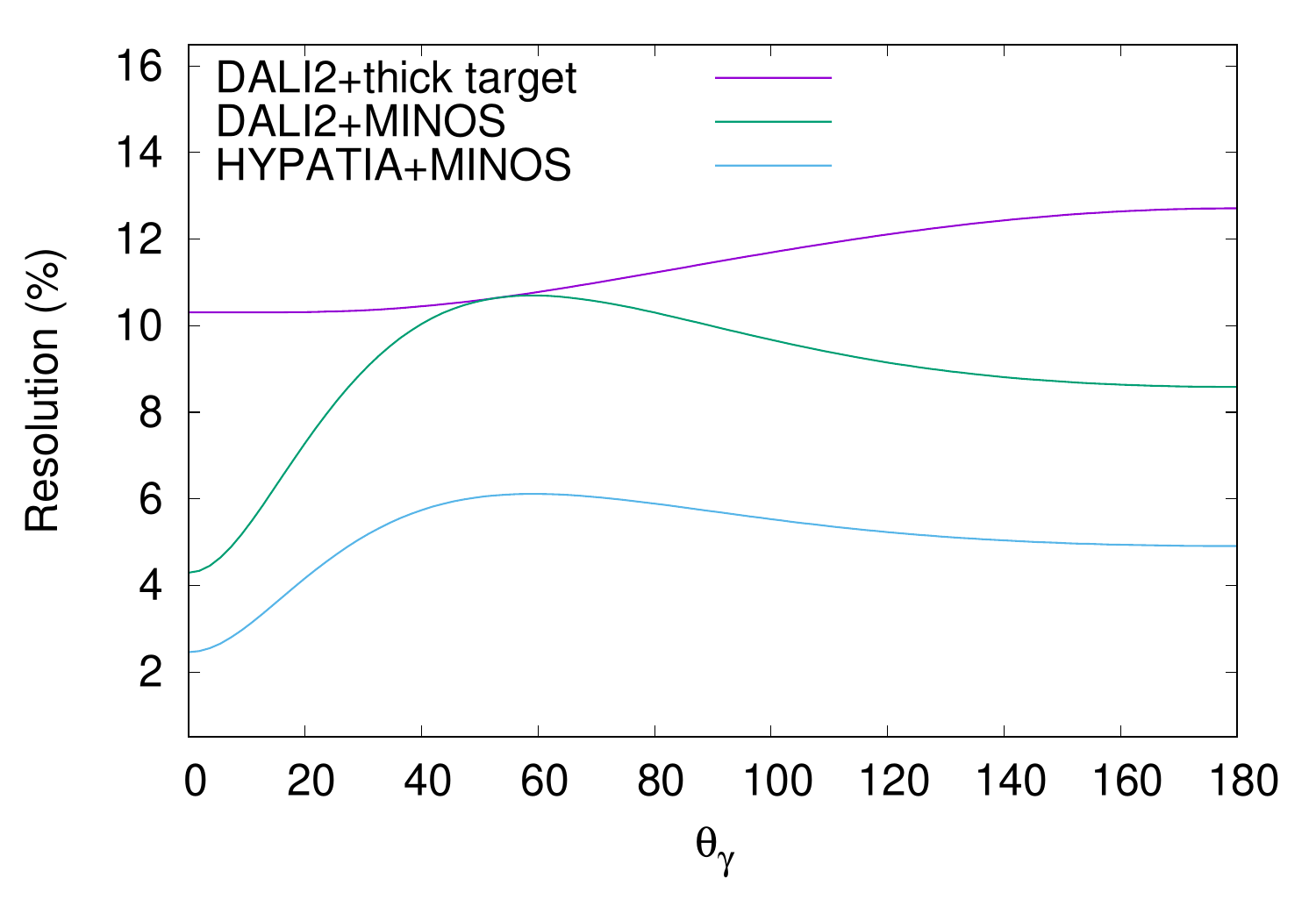}
    \caption{Total energy resolution (FWHM) for different \gammaray spectrometers.  For a thick target, a large  $\Delta\beta$ is obtained. This uncertainty is reduced by using MINOS. The use of smaller detectors with better intrinsic resolution, as the HYPATIA array, further improve the achievable energy resolution. The calculations were performed for $\beta=0.6$ and  a 1~MeV \gammaray energy assuming a square-root dependence of the intrinsic energy resolution.}
    \label{fig:deltaE}
\end{figure}

Observed \gammaray energies, $E_\gamma$, are Doppler shifted with respect to the emitted energies in the rest frame, $E_{\gamma 0}$, following the relation
\begin{equation}
    \frac{E_{\gamma0}}{E_{\gamma}} = \frac{1-\beta\cos\theta_\gamma}{\sqrt{1-\beta^2}}, \label{eq:Doppler}
\end{equation}
where $\beta$ denotes the velocity of the beam and $\theta_\gamma$ is the angle between the particle direction and the $\gamma$-ray emission in the laboratory frame of reference. The energy resolution that can be achieved is therefore limited by the uncertainties in the velocity, $\Delta\beta$, the emission angle of the \gammaray, $\Delta\theta_\gamma$, and the intrinsic resolution of the detection material, $\Delta E_\gamma$. 
In general, the use of thicker secondary targets leads to a higher energy loss and therefore a larger contribution to $\Delta\beta$. The use of  MINOS~\cite{Obertelli_EPJA_2014} overcomes this problem  by tracking the protons ejected from the target after knockout reactions and reconstructing the interaction position. In this way thicker targets can be used without affecting the achievable energy resolution as shown in Fig.~\ref{fig:deltaE}.
While the intrinsic resolution, $\Delta E_\gamma$, is a property of the detector material, 
the angular resolution, $\Delta\theta_\gamma$, can be improved  either by reducing the size of the detectors or by implementing a position-sensitive readout with a sub-centimeter resolution for the first interaction of the \gammarays. 

In addition to the energy resolution, the total efficiency of the detection array is an important factor. The use of high-efficiency materials becomes critical for the study of nuclei at the limit of existence where few events are expected, and for reconstructing complicated level schemes with  multiple \gammaray emissions. In general, \gammaray detection arrays can be optimized either for an increased energy resolution or for a high detection efficiency. 

A further parameter to considered is the time resolution. The high background generated in these type of experiments is usually a limiting factors, specially at low energies. It can be reduced by  selecting events correlated to the impinging particle within a short time window. The successful use of this technique requires a time resolution typically better than 1~ns. Such resolution is only achievable for some scintillation materials.

The interplay of different factors makes it necessary to tackle the next generation \gammaray arrays from different perspectives, which depend on the physics cases pursued. In the following, we make a summary of future projects regarding \gammaray detection arrays at the RIBF.

\subsection{Towards a new scintillation detector array: The HYPATIA project} 

Since 2001 the main device used at the RIBF to perform in-beam $\gamma$-ray spectroscopy has been the DALI2 array~\cite{takeuchi2014dali2}, composed of 186 NaI(Tl) detectors.
From 2017 the upgraded DALI2$^+$ array~\cite{Murray_RAPR_2018}, consisting of 226 NaI(Tl) detectors, covering a large solid angle and offering a high detection efficiency, has been employed to detect radiation emitted primarily after direct reactions with RIBs. 

Significant results have been obtained using the DALI2 and DALI2$^+$ array, notably the discovery of a new magic number at $N=34$~\cite{Steppenbeck_Nature_2013}, the spectroscopy of first excited states in the \say{Island of Inversion}~\cite{Doornenbal_PhysRevLett.103.032501, Doornenbal_PhysRevLett.111.212502}, and the first spectroscopy of the most exotic neutron-rich isotopes accessible to date within the SEASTAR project~\cite{seastar_web}. Given the future upgrade of the RIBF~\cite{ribf_upgrade}, a new $\gamma$-ray detection array is required to fully exploit the new experimental capabilities that come along with it. 

\begin{figure}[htb]
\begin{center}
    \includegraphics[width=0.48\textwidth]{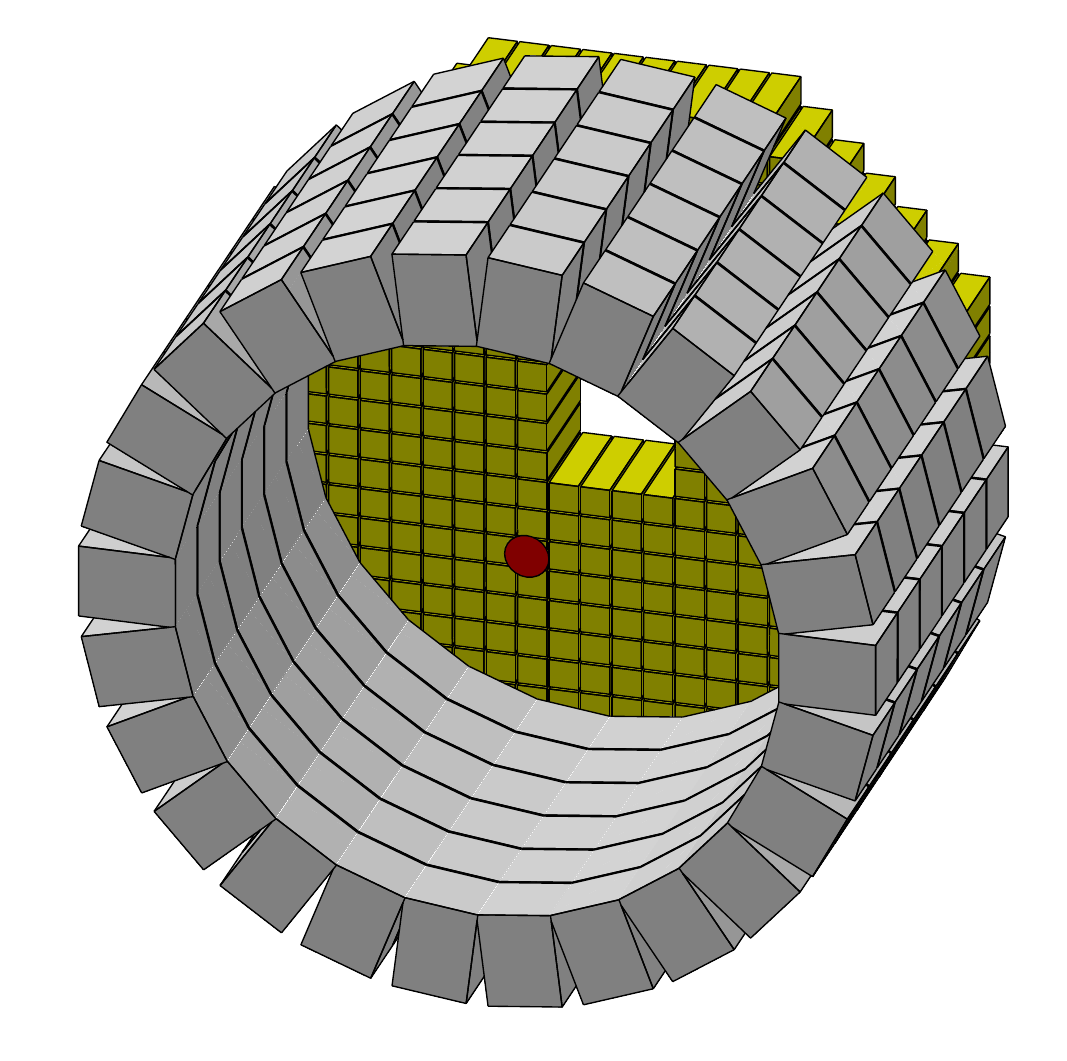}
    \end{center}
    \caption{Conceptual design of the HYPATIA array composed of a barrel of CeBr$_3$ detectors and a forward wall of HR-GAGG crystals.}
    \label{fig:hypatia}
\end{figure}

Based on the experience acquired during the last two decades using DALI2 and DALI2$^+$, the key characteristics desired for this new array are:
\begin{itemize}
    \item Higher detection efficiency which would allow to reach more exotic isotopes, as well as to perform detailed $\gamma-\gamma$ coincidences analysis to build level schemes.
    \item Better energy resolution to identify previously unresolved transitions and exploit lifetime effects.
    \item Improved time resolution to suppress the beam-related background.
    \item A versatile design which allows to install of the array at different locations and with varying configurations optimized for each measurement.
\end{itemize}
With this requirement in mind, the HYPATIA (HYbrid Photon detector Array To Investigate Atomic nuclei) array has been designed~\cite{hypatia_web}.  This array, depicted in Fig.~\ref{fig:hypatia}, consists of 624 Cerium Bromide (CeBr$_3$) detectors of $30\times 30\times 80$~mm$^2$ forming a barrel around the target and 384 Cerium-doped Gadolinium Aluminum Gallium Garnet (HR-GAGG) detectors of $25\times 25\times 100$~mm$^2$ placed as a forward wall.  The CeBr$_3$ detectors were chosen due to their improved energy and time resolution compared to NaI(Tl), their reasonable cost, and the lack of intrinsic radiation. On the other hand, the HR-GAGG crystals are non-hygroscopic and have a high effective atomic number, therefore offering the possibility to minimize the dead space and increase the efficiency and peak-to-total ratio at forward angles. 
Thanks to the cuboidal shape of all the detectors, it is possible to arrange them in different configurations, including surrounding the MINOS or STRASSE targets, as well as to place the array at the different spectrometers of the RIBF. Furthermore, the selected shapes and sizes of the detectors are compatible with the mechanical structure of the DALI2$^+$ array, therefore it will be possible to gradually replace the detectors without interrupting the physics program. 

The future availability of $^{238}$U primary beam at 345~MeV/nucleon with an intensity of 2000~pnA would allow the production of very exotic isotopes towards the drip lines. Inelastic scattering, Coulomb excitation, transfer reaction, and quasi-free scattering reactions can be employed to study these isotopes. Thanks to its superior efficiency, time resolution, and peak-to-total ratio, the HYPATIA array will be able to overcome some of the limitations of DALI2$^+$, as well as of high-resolution Ge detectors which do not cover a solid angle of $4\pi$, such as the pandemonium effect. A key example of the possibilities of HYPATIA with the fast beam at the RIBF is the first spectroscopy of $^{60}$Ca.

\begin{figure}[bth]
\begin{center}
  \includegraphics[width=\textwidth]{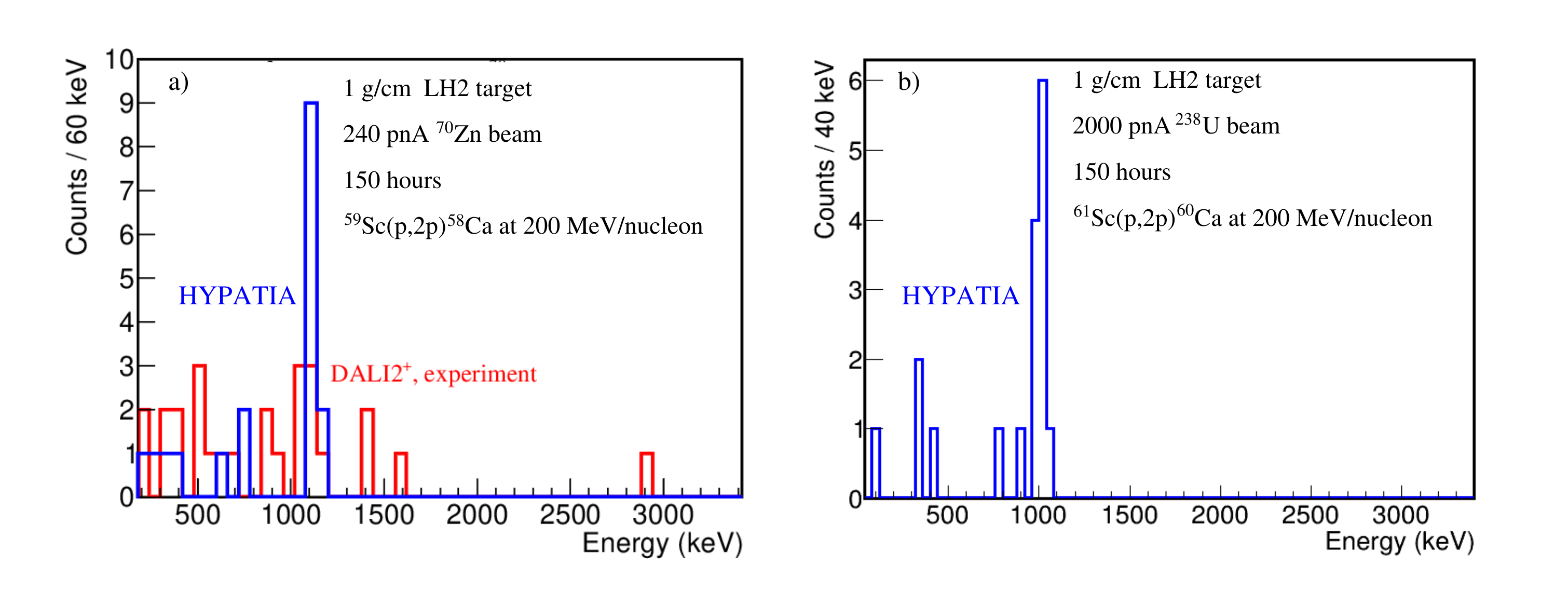}
    \end{center}
    \caption{a) Comparison between the performance of DALI2$^+$ and HYPATIA for the measurement of $^{58}$Ca. b)~Expected Doppler-corrected $\gamma$ spectrum of $^{60}$Ca for a 150-hours measurement with  HYPATIA.}
    \label{fig:ca60_hypatia}
\end{figure}

\noindent
The Ca isotopic chain is particularly interesting for nuclear structure studies. While the proton shell closure at $Z=20$  seems to be maintained, new neutron shell closures at $N=32$ and $N=34$ have been evidenced by spectroscopy~\cite{Huck_PhysRevC.31.2226,Steppenbeck_Nature_2013}, mass measurements~\cite{Wienholtz_Nature2013,Michimasa_PhysRevLett.121.022506}, and knockout reaction cross sections~\cite{Gade_PhysRevC.74.021302,Chen_PhysRevLett.123.142501}.
The possibility of a further magic number at the $N=40$ harmonic-oscillator shell closure has been widely studied~\cite{Gade_physics3040077, Brown_physics4020035}.  Although measurements on $^{58}$Ca~\cite{Chen_PLB_2023138025} and $^{62}$Ti~\cite{Cortes_PLB_2020135071}, the closest neighbors of $^{60}$Ca, show no indication of a shell closure at $N=40$, only the first spectroscopy of $^{60}$Ca can provide definitive answers. This measurement will be possible with the HYPATIA array, as shown in Fig.~\ref{fig:ca60_hypatia}b),  where the simulated $\gamma$-ray spectrum obtained in a 150~h measurement is shown. As a reference, Fig.~\ref{fig:ca60_hypatia}a) shows a comparison between the experimental spectrum obtained for $^{58}$Ca measured with DALI2$^+$~\cite{Chen_PLB_2023138025}, and the expected performance of HYPATIA under the same conditions.

\subsection{Future projects with Ge detectors}

The \gammaray tracking technique of Ge detectors is the state-of-the-art method to achieve the best energy resolution for \gammarays. This technique has been implemented in the latest large-scale Ge arrays GRETINA~\cite{PASCHALIS201344} and  AGATA~\cite{AKKOYUN201226}. 
During 2020 and 2021, the RIBF hosted the HiCARI array~\cite{Wimmer2021} to conduct the first studies with tracking-type germanium detectors, which yielded successful outcomes. HiCARI was organized as a one-time campaign of experiments at the RIBF by gathering Ge detectors from other facilities.
It is desirable to make continued efforts to push in-beam \gammaray spectroscopy based on tracking technology forward. For this reason, the Gamma-ray Tracking in R5 (GT-5) project was launched in 2023 at the RIKEN Nishina Center as a common and long-term platform for tracking-type Ge detectors at the RIBF to drive physics experiments and instrumentation for tracking technologies. 
The GT-5 has its own scientific and development programs related to in-beam \gammaray spectroscopy using RIBs. On a hardware basis, the GT-5 presently encompasses two in-house tracking detectors. One is the Compton camera telescope GREI~\cite{motomura_IEEE_2007}, and the other is a position-sensitive strip germanium telescope SGT~\cite{Suzuki_RAPR}, both based on planar crystals with stripped segments. 
Furthermore, a GRETINA quad-type module was procured in 2023. In addition to the detectors, a maintenance facility for Ge detectors, electronics, and infrastructure is arranged.
A unique aspect of the GT-5 is that it is not only involved in the scientific activities of the RIBF but also in the framework of the TRIP (Transformative Research Innovation Platform of RIKEN platforms) project, a RIKEN-wide initiative starting in 2023. 
One of the main goals of the TRIP use case is to establish an optical model for radioactive isotopes that are needed for reaction analyses and predictions but have yet to be studied based on systematic experimental data. To provide systematic data to optimize optical model potential parameters, the GT-5 is teamed with the elastic scattering measurement program MESA (Measurement of Elastic Scattering Anytime anywhere on any beam) and the reaction cross-section measurement program S3CAN (Symbiotic Systematic and Simultaneous Cross-section measurements for All over the Nuclear chart). As the first opportunities for data collection, experiments with stable beams and with RI beams have been conducted in 2024.

\section{Neutron detection for invariant-mass spectroscopy}\label{sec:neutrons}
Invariant mass spectroscopy is a powerful tool for measuring the properties of nuclei under extreme isospin conditions. 
In particular, for neutron-rich isotopes, phenomena such as halo structures near the neutron drip line, neutron correlations, unbound states, and new cluster states are to be expected. The study of such properties allows us not only to probe the mechanisms underlying the nuclear force but also to better understand the nuclear equation of state and to investigate the properties of neutron stars.
Such neutron-rich systems can be produced using different types of reactions with radioactive beams, and their structure can be determined by measuring the momenta of all the outgoing particles, including neutrons, following the decay of any intermediate resonance.

Kinematically complete measurements are performed at the SAMURAI spectrometer of the RIBF. For the detection of the fast neutrons evaporated after the reactions, the NEBULA (NEutron detection system for Breakup of Unstable nuclei with Large Acceptance) detector has been employed~\cite{Kondo_NIM_2020}.
NEBULA is made up of two identical wall structures, which consist of a layer of VETO scintillation detectors to eliminate hits from charged particles and two-layered neutron detectors with 30 each scintillation bars with readouts by PMTs at both ends. Each neutron detection bar is shaped as $12\times12\times180~\textrm{cm}^3$, featuring high neutron detection efficiency, large angular coverage, and capability of multiple neutron coincidence detection.
A further improvement in the multiple neutron detection efficiency was achieved by combining NEBULA with NeuLAND (New Large-Area Neutron Detector), a new neutron detector array developed for the R3B setup under construction at FAIR~\cite{Boretzky_NIM_2021}. 
With this combined setup, a diversity of phenomena has been measured, such as the first measurement of $^{28}$O~\cite{Kondo_Nature_2023}.

The upgrade plans of the RIBF will enable reaching of further neutron-rich isotopes, which are expected to have various weakly bound multi-neutron systems.  To extend the sensitivity of multi-neutron reconstructions, a project to extend the detector volume for better detection efficiency, NEBULA Plus, has recently been completed through a collaborative effort between Japan and LPC Caen in France, as part of the French-funded EXPAND project~\cite{expand}. The NEBULA Plus array consisting of 90 new modules has been integrated with the existing array, effectively doubling the number of detection layers. This expansion increases the one-neutron detection efficiency by nearly a factor of 2 and improves the four-neutron detection efficiency by almost one order of magnitude.
Initial in-beam tests were carried out at the RIBF in 2024~\cite{NebulaPlus_RAPR}. Physics opportunities such as the investigation of the two-neutron decay of $^{21}$B, the search for $^{21}$B$^*$, and the measurement of a three-neutron decay of $^{20}$B, are expected to be accessible with the NEBULA Plus detector.

In contrast, a detector focused on more precise measurement by improving the achievable resolution in measurements of neutron momentum is also being constructed. The High-resolution Detector Array for Multi-neutron Events (HIME) detector was originally developed by the group of The Tokyo Institute of Technology (now Institute of Science Tokyo), and further developed in collaboration with the Technical University of Darmstadt. The optimized energy resolution of HIME is achieved through the use of smaller scintillator bars with dimensions of 4 $\times$ 2 $\times$ 100~cm$^3$, which improves position resolution, and new electronics for a higher time precision. It consists of layers of scintillator bars arranged alternating in $x$ and $y$ directions~\cite{Knoesel_thesis}. This arrangement will not only improve its resolution, but will also allow discriminating the hits due to recoiling protons, allowing the selection of valid events and reducing cross-talk~\cite{Hime_report}.
Currently, a prototype of HIME has been built, and more layers are under construction at TU Darmstadt. The improved energy resolution will enable precise measurements, such as the neutron-neutron scattering length, which characterizes the two-neutron interactions at low energies.
Besides that, it is planned to be used for the detection of the recoil neutrons after the $(p,pn)$ reactions in combination with the proton trackers as described in the previous sections.

\section{Summary}\label{sec:conclusion}

Thanks to advances on the production of RIBs, isotopes that were previously inaccessible have been studied, and many interesting phenomena have been discovered. 
An integral part of research on the structure of exotic isotopes is the use of appropriate devices to detect the \gammarays, charged particles, and neutrons produced in the reactions.

A successful campaign of QFS experiments was performed at the RIBF using the MINOS LH$_2$ target. The detection devices used for these experiments provided a plethora of results, and highlighted the limitations and possible improvements of the different detection systems. This experience, together with the future upgrade of the RIBF have promoted the development of new and improved detectors. 

Based on the good performance of MINOS, the STRASSE Si tracker is under development. In addition to an improved resolution on the vertex reconstruction for knockout reactions, this system will provide the possibility to perform missing-mass measurement by the coupling with CATANA PLUS. 
A similar development, TOGAXI, will enable dedicated measurements of cluster knockout reaction. Thanks to the use of a calorimeter, missing mass-measurements will also be possible. 
Invariant mass measurements will benefit from the development of the neutron detectors NEBULA Plus and HIME, optimized for an increased efficiency and resolution, respectively. 
Furthermore, the study of bound excited states will profit from the developments on \gammaray detections arrays, such as HYPATIA, which will provide a better energy and time resolutions, as well as an improved peak-to-total ratio. Applications requiring optimized energy resolution can be addressed within the framework of the GT5 project. 

In the future, the combination of QFS experiments with these new detection devices will help us to broaden our understanding of the structure of exotic isotopes. These detectors will also provide benchmarks for future developments in combination with the upgraded intesities of the RIBF. 

\section*{Acknowledgment}
The authors acknowledge the contributions of Pieter Doornenbal, Marco Kn\"{o}sel, Alexandre Obertelli, Nigel Orr, Valerii Panin, Marina Petri, Daisuke Suzuki, Yasuhiro Togano, Ryotaro Tsuji, Tomohiro Uesaka, and Rin Yokoyama.
J.T. acknowledges the ONOKORO collaboration and the support from the Grants-in-Aid for Scientific Research of the Japan Society for the Promotion of Science (JSPS) KAKENHI, Grant No. JP21H04975. M.L.C acknowledges the support of the RIKEN Special Postdoctoral Researcher Program. H.N.L thanks the STRASSE-CATANA collaboration, and acknowledges the support from the National Key R\&D Program of China (Grant No. 2023YFA1606403, No. 2024YFE0102800), the National Natural Science Foundation of China (Grant No. 12375111), and the Fundamental Research Funds for the Central Universities of China.
R.T acknowledges the support from the UK STFC and the Royal Society.




\vspace{0.2cm}
\noindent


\let\doi\relax


\begin{thebibliography}{68}

\bibitem{Nowacki_PPNP_120_2021}
F.~Nowacki, A.~Obertelli, and A.~Poves, Prog. Part. Nucl. Phys., {\bf 120}, 103866 (2021).

\bibitem{Aumann_PPNP_112_2020}
T.~Aumann and C.~A. Bertulani, Prog. Part. Nucl. Phys., {\bf 112}, 103753 (2020).

\bibitem{Yano_NIMB2007}
Y.~Yano, Nucl. Instrum. Methods Phys. Res. B, {\bf 261}, 1009 (2007).

\bibitem{Accelerator2021_RAPR}
T.~Nakamura, K.~Ozeki, T.~Dantsuka, et~al., RIKEN Accel. Prog. Rep 2021, {\bf 55}, 177 (2022).

\bibitem{Accelerator2022_RAPR}
K.~Kobayashi, K.~Suda, T.~Adachi, et~al., RIKEN Accel. Prog. Rep. 2022, {\bf 56}, 207 (2023).

\bibitem{Kondo_Nature_2023}
Y.~Kondo, N.~L. Achouri, H.~Al Falou, et~al., Nature, {\bf 620}(7976), 965 (2023).

\bibitem{Ahn_PhysRevLett.129.212502_2022}
D.~S. Ahn, J.~Amano, H.~Baba, et~al., Phys. Rev. Lett., {\bf 129}, 212502 (2022).

\bibitem{Steppenbeck_Nature_2013}
D.~Steppenbeck, S.~Takeuchi, N.~Aoi, et~al., Nature, {\bf 502}(7470), 207 (2013).

\bibitem{Taniuchi_Nature_2019}
R.~Taniuchi, C.~Santamaria, P.~Doornenbal, et~al., Nature, {\bf 569}(7754), 53 (2019).

\bibitem{Kisamori_PhysRevLett_116_2016}
K.~Kisamori, S.~Shimoura, H.~Miya, et~al., Phys. Rev. Lett., {\bf 116}, 052501 (2016).

\bibitem{Duer_Nature_2022}
M.~Duer, T.~Aumann, R.~Gernh{\"a}user, et~al., Nature, {\bf 606}(7915), 678 (2022).

\bibitem{Kubo_PTEP_2012}
T.~Kubo, D.~Kameda, H.~Suzuki, et~al., Prog. Theor. Exp. Phys., {\bf 2012}(1), 03C003 (2012).

\bibitem{Fukuda_NIMA2013}
N.~Fukuda, T.~Kubo, T.~Ohnishi, N.~Inabe, H.~Takeda, D.~Kameda, and H.~Suzuki, Nucl. Instrum. Methods Phys. Res. B, {\bf 317}, 323 (2013).

\bibitem{Kobayashi_NIM_317_2013}
T.~Kobayashi, N.~Chiga, T.~Isobe, et~al., Nucl. Instrum. Methods Phys. Res. B, {\bf 317}, 294 (2013).

\bibitem{Michimasa_NIMA_317_2013}
S.~Michimasa, M.~Takaki, Y.~Sasamoto, et~al., Nucl. Instrum. Methods Phys. Res. B, {\bf 317}, 305 (2013).

\bibitem{Michimasa_PTEP_2019}
S.~Michimasa, J.~Hwang, K.~Yamada, et~al., Prog. Theor. Exp. Phys., {\bf 2019}(4), 043D01 (2019).

\bibitem{YAMAGUCHI2013}
Y.~Yamaguchi, M.~Wakasugi, T.~Uesaka, et~al., Nucl. Instrum. Methods Phys. Res. B, {\bf 317}, 629 (2013).

\bibitem{Obertelli_EPJA_2014}
A.~Obertelli, A.~Delbart, S.~Anvar, et~al., Eur. Phys. J. A, {\bf 50}(1), 8 (2014).

\bibitem{santamaria2018tracking}
C.~Santamaria, A.~Obertelli, S.~Ota, et~al., Nucl. Instrum. Methods Phys. Res. A, {\bf 905}, 138 (2018).

\bibitem{takeuchi2014dali2}
S.~Takeuchi, T.~Motobayashi, Y.~Togano, M.~Matsushita, N.~Aoi, K.~Demichi, H.~Hasegawa, and H.~Murakami, Nucl. Instrum. Methods Phys. Res. A, {\bf 763}, 596 (2014).

\bibitem{Murray_RAPR_2018}
I.~Murray, F.~Browne, S.~Chen, et~al., RIKEN Accel. Prog. Rep., {\bf 51}, 158 (2017).

\bibitem{Boretzky_NIM_2021}
K.~Boretzky, I.~Ga\v{s}pari\'{c}, M.~Heil, et~al., Nucl. Instrum. Methods Phys. Res. A, {\bf 1014}, 165701 (2021).

\bibitem{Kondo_NIM_2020}
Y.~Kondo, T.~Tomai, and T.~Nakamura, Nucl. Instrum. Methods Phys. Res. B, {\bf 463}, 173 (2020).

\bibitem{Taniuchi_PTEP_2025}
R.~Taniuchi, Prog. Theor. Exp. Phys., page ptaf084 (2025).

\bibitem{Lui_PTEP_2025}
H.~Liu, S.~Chen, and F.~Browne, Prog. Theor. Exp. Phys. (2025).

\bibitem{Cortes_PTEP_2025}
M.~L. Cort\'{e}s, Prog. Theor. Exp. Phys., page ptaf062 (2025).

\bibitem{Kahlbow_PTEP_2025}
J.~Kahlbow, Prog. Theor. Exp. Phys. (2025).

\bibitem{Chen_PTEP_2025}
S.~Chen, F.~Browne, T.~Rodr\'{i}guez, and V.~Werner, Prog. Theor. Exp. Phys. (2025).

\bibitem{ribf_upgrade}
{RIBF} {F}acility {U}pgrade {P}roject,
\newblock \url{https://www.nishina.riken.jp/researcher/RIBFupgrade/index_e.html} (2024).

\bibitem{ermisch2005systematic}
K.~Ermisch, H.~R. Amir-Ahmadi, A.M. Van Den~Berg, et~al., Phys. Rev. C, {\bf 71}(6), 064004 (2005).

\bibitem{hasell1986elastic}
D.~K. Hasell, A.~Bracco, H.~P. Gubler, et~al., Phys. Rev. C, {\bf 34}(1), 236 (1986).

\bibitem{moss1980proton}
G.~A. Moss, L.~G. Greeniaus, J.~M. Cameron, et~al., Phys. Rev. C, {\bf 21}(5), 1932 (1980).

\bibitem{strasse_epja}
H.~N. Liu, F.~Flavigny, H.~Baba, et~al., Eur. Phys. J. A, {\bf 59}(6), 121 (2023).

\bibitem{catana}
Y.~Togano, T.~Nakamura, Y.~Kondo, et~al., Nucl. Instrum. Methods Phys. Res. B, {\bf 463}, 195 (2020).

\bibitem{prespec}
C.~Louchart, J.M. Gheller, Ph. Chesny, et~al., Nucl. Instrum. Methods Phys. Res. A, {\bf 736}, 81 (2014).

\bibitem{sts-xyter}
K.~Kasinski, A.~Rodriguez-Rodriguez, J.~Lehnert, W.~Zubrzycka, R.~Szczygiel, P.~Otfinowski, R.~Kleczek, and C.J. Schmidt, Nucl. Instrum. Methods Phys. Res. A, {\bf 908}, 225 (2018).

\bibitem{Tanaka_Science_2021}
J.~Tanaka, Z.~Yang, S.~Typel, et~al., Science, {\bf 371}(6526), 260 (2021).

\bibitem{matsuda2011large}
Y.~Matsuda, H.~Sakaguchi, J.~Zenihiro, et~al., Nucl. Instrum. Methods Phys. Res. A, {\bf 643}(1), 6 (2011).

\bibitem{chebotaryov2018proton}
S.~Chebotaryov, S.~Sakaguchi, T.~Uesaka, et~al., Prog. Theor. Exp. Phys., {\bf 2018}(5), 053D01 (2018).

\bibitem{matsuda2013elastic}
Y.~Matsuda, H.~Sakaguchi, H.~Takeda, et~al., Phys. Rev. C, {\bf 87}(3), 034614 (2013).

\bibitem{verde2013farcos}
G.~Verde, L.~Acosta, T.~Minniti, et~al., J. Phys. Conf. Ser., {\bf 420}, 012158 (2013).

\bibitem{li2023validation}
P.~J. Li, D.~Beaumel, J.~Lee, et~al., Phys. Rev. Lett., {\bf 131}(21), 212501 (2023).

\bibitem{TANAKA20234}
J.~Tanaka, R.~Tsuji, K.~Higuchi, et~al., Nucl. Instrum. Methods Phys. Res. B, {\bf 542}, 4 (2023).

\bibitem{uesaka2021comprehensive}
T.~Uesaka, J.~Zenihiro, and K.~Ogata,
\newblock Comprehensive research of cluster formation in nuclear matter,
\newblock Grant-in-Aid for Specially Promoted Research, JSPS (2021).

\bibitem{HIGUCHI202384}
K.~Higuchi, J.~Tanaka, R.~Tsuji, et~al., Nucl. Instrum. Methods Phys. Res. B, {\bf 542}, 84 (2023).

\bibitem{Doornenbal_PTEP_2012}
P.~Doornenbal, Prog. Theor. Exp. Phys., {\bf 2012}(1), 03C004 (12 2012).

\bibitem{Doornenbal_PhysRevLett.103.032501}
P.~Doornenbal, H.~Scheit, N.~Aoi, et~al., Phys. Rev. Lett., {\bf 103}, 032501 (2009).

\bibitem{Doornenbal_PhysRevLett.111.212502}
P.~Doornenbal, H.~Scheit, S.~Takeuchi, et~al., Phys. Rev. Lett., {\bf 111}, 212502 (2013).

\bibitem{seastar_web}
The {SEASTAR} project,
\newblock \url{https://www.nishina.riken.jp/collaboration/SUNFLOWER/experiment/seastar/index.php} (2024).

\bibitem{hypatia_web}
The {HYPATIA} array,
\newblock \url{https://www.nishina.riken.jp/collaboration/SUNFLOWER/devices/hypatia/index.php} (2024).

\bibitem{Huck_PhysRevC.31.2226}
A.~Huck, G.~Klotz, A.~Knipper, C.~Mieh\'e, C.~Richard-Serre, G.~Walter, A.~Poves, H.~L. Ravn, and G.~Marguier, Phys. Rev. C, {\bf 31}, 2226 (1985).

\bibitem{Wienholtz_Nature2013}
F.~Wienholtz, D.~Beck, K.~Blaum, et~al., Nature, {\bf 498}(7454), 346 (2013).

\bibitem{Michimasa_PhysRevLett.121.022506}
S.~Michimasa, M.~Kobayashi, Y.~Kiyokawa, et~al., Phys. Rev. Lett., {\bf 121}, 022506 (2018).

\bibitem{Gade_PhysRevC.74.021302}
A.~Gade, R.~V.~F. Janssens, D.~Bazin, et~al., Phys. Rev. C, {\bf 74}, 021302 (2006).

\bibitem{Chen_PhysRevLett.123.142501}
S.~Chen, J.~Lee, P.~Doornenbal, et~al., Phys. Rev. Lett., {\bf 123}, 142501 (2019).

\bibitem{Gade_physics3040077}
A.~Gade, Physics, {\bf 3}(4), 1226--1236 (2021).

\bibitem{Brown_physics4020035}
B.~Alex Brown, Physics, {\bf 4}(2), 525--547 (2022).

\bibitem{Chen_PLB_2023138025}
S.~Chen, F.~Browne, P.~Doornenbal, et~al., Phys. Lett. B, {\bf 843}, 138025 (2023).

\bibitem{Cortes_PLB_2020135071}
M.~L. Cort\'{e}s, W.~Rodriguez, P.~Doornenbal, et~al., Phys. Lett. B, {\bf 800}, 135071 (2020).

\bibitem{PASCHALIS201344}
S.~Paschalis, I.Y. Lee, A.O. Macchiavelli, et~al., Nucl. Instrum. Methods Phys. Res. A, {\bf 709}, 44 (2013).

\bibitem{AKKOYUN201226}
S.~Akkoyun, A.~Algora, B.~Alikhani, et~al., Nucl. Instrum. Methods Phys. Res. A, {\bf 668}, 26 (2012).

\bibitem{Wimmer2021}
K.~Wimmer, P.~Doornenbal, N.~Aoi, et~al., RIKEN Accel. Prog. Rep. 2020, {\bf 54}, S27 (2021).

\bibitem{motomura_IEEE_2007}
S.~Motomura, S.~Enomoto, H.~Haba, K.~Igarashi, Y.~Gono, and Y.~Yano, IEEE Trans. Nucl. Sci., {\bf 54}(3), 710 (2007).

\bibitem{Suzuki_RAPR}
M.~K. Suzuki, T.~K. Onishi, H.~J. Ong, H.~Suzuki, Y.~Ichikaw, N.~Imai, N.~Aoi, H.~Iwasaki, and H.~Sakurai, RIKEN Accel. Prog. Rep. 2003, {\bf 37}, 153 (2024).

\bibitem{expand}
The {EXPAND} project,
\newblock \url{https://www.in2p3.cnrs.fr/sites/institut_in2p3/files/page/2019-07/7-Doc-ORR.pdf} (2019).

\bibitem{NebulaPlus_RAPR}
Y.~Kondo N.~A.~Orr, J.~Gibelin and H.~Otsu for~the NEBULA-Plus~collaboration, RIKEN Accel. Prog. Rep. 2022, {\bf 56}, 91 (2023).

\bibitem{Knoesel_thesis}
M.~Kn\"{o}sel,
\newblock {\em Measurement of the Neutron-Neutron Scattering Length},
\newblock PhD thesis, Technische Universitaet Darmstadt (2023).

\bibitem{Hime_report}
T.~Nakamura and R.~Tanaka for~the NEBULA Plus~collaboration,
\newblock Development of a next-generation neutron detector array {HIME},
\newblock {\url{http://be.nucl.ap.titech.ac.jp/~ryuki/report/development_of_hime.pdf}} ().

\end{thebibliography}

\end{document}